\documentclass[acmsmall]{acmart}

\AtBeginDocument{%
  \providecommand\BibTeX{{%
    \normalfont B\kern-0.5em{\scshape i\kern-0.25em b}\kern-0.8em\TeX}}}

\copyrightyear{2026}
\acmYear{2026}
\setcopyright{cc}
\setcctype{by}
\acmConference[CSCW Companion '26]{Companion of the Computer-Supported Cooperative Work and Social Computing}{October 10--14, 2026}{Salt Lake City, UT, USA}
\acmBooktitle{Companion of the Computer-Supported Cooperative Work and Social Computing (CSCW Companion '26), October 10--14, 2026, Salt Lake City, UT, USA}
\acmDOI{10.1145/3785651.3831515}
\acmISBN{979-8-4007-2378-0/2026/10}

\makeatletter
\AtBeginDocument{%
  \fancypagestyle{standardpagestyle}{%
    \fancyhf{}%
    \renewcommand{\headrulewidth}{\z@}%
    \renewcommand{\footrulewidth}{\z@}%
    \fancyhead[LE]{\sffamily\footnotesize\thepage}%
    \fancyhead[RO]{\sffamily\footnotesize\thepage}%
    \fancyhead[RE]{\sffamily\footnotesize\@shortauthors}%
    \fancyhead[LO]{\sffamily\footnotesize\shorttitle}}%
  \fancypagestyle{firstpagestyle}{%
    \fancyhf{}%
    \renewcommand{\headrulewidth}{\z@}%
    \renewcommand{\footrulewidth}{\z@}}%
  \pagestyle{standardpagestyle}%
}
\makeatother

\begin{document}

\title[Frontstage Mediation Work]{Frontstage Mediation Work: Invisible Work Bridging Gaps Between AI Decisions and User Expectations}
\author{Yongjae Sohn}
\affiliation{%
  \institution{KAIST}
  \department{Department of Industrial Design}
  \city{Daejeon}
  \country{Republic of Korea}
}
\email{yongjae.sohn@kaist.ac.kr}

\author{Daehyun Kwak}
\affiliation{%
  \institution{KAIST}
  \department{Department of Industrial Design}
  \city{Daejeon}
  \country{Republic of Korea}
}
\email{dh.kwak@kaist.ac.kr}

\author{Jiyeon Amy Seo}
\affiliation{%
  \institution{University of Michigan}
  \department{School of Information}
  \city{Ann Arbor, Michigan}
  \country{United States}
}
\email{amyeony@umich.edu}

\author{Hyungjun Cho}
\affiliation{%
  \institution{University of Florida}
  \department{School of Art + Art History}
  \city{Gainesville, Florida}
  \country{United States}
}
\email{hyungjuncho@ufl.edu}

\author{Seongah Youn}
\affiliation{%
  \institution{Seoul National University}
  \department{Design}
  \city{Seoul}
  \country{Republic of Korea}
}
\email{seongah7865@snu.ac.kr}

\author{Youn-kyung Lim}
\affiliation{%
  \institution{KAIST}
  \department{Department of Industrial Design}
  \city{Daejeon}
  \country{Republic of Korea}
}
\email{younlim@kaist.ac.kr}
\renewcommand{\shortauthors}{Sohn et al.}

\begin{abstract}
Automated service systems increasingly generate algorithmic operational decisions that shape how services are delivered. However, these decisions reach end-users only through frontline workers who carry them out in real-world settings. During this process, automated decisions can diverge from user expectations, surfacing as friction at the service encounter. We propose Frontstage Mediation Work as a preliminary analytic lens for examining the often invisible labor through which frontline workers anticipate and manage such misalignments between algorithmic decisions and user expectations. Drawing on a qualitative case study of an On-Demand Ride-Pooling service, we identify four recurring practices through which drivers sustain the service encounter when frictions arise. Such labor remains absorbed into routine operations, leaving no trace in performance metrics, system logs, or formal job descriptions. This paper contributes to worker-centered HCI scholarship by illustrating how automated services shift onto frontline workers the responsibility of managing the interactional consequences of system-level decisions.
\end{abstract}

\begin{CCSXML}
<ccs2012>
<concept>
<concept_id>10003120.10003130.10003131</concept_id>
<concept_desc>Human-centered computing~Collaborative and social computing theory, concepts and paradigms</concept_desc>
<concept_significance>500</concept_significance>
</concept>
<concept>
<concept_id>10003120.10003121.10011748</concept_id>
<concept_desc>Human-centered computing~Empirical studies in HCI</concept_desc>
<concept_significance>500</concept_significance>
</concept>
</ccs2012>
\end{CCSXML}

\ccsdesc[500]{Human-centered computing~Collaborative and social computing theory, concepts and paradigms}
\ccsdesc[500]{Human-centered computing~Empirical studies in HCI}

\keywords{Frontstage mediation work; Invisible labor; Algorithmic management; Emotional labor; Articulation work; Frontline workers; On-demand ride-pooling; Worker-centered design}

\maketitle

\section{INTRODUCTION AND CONCEPTUAL BACKGROUND}
Automated systems, including AI, are increasingly deployed in service operations with promises of efficiency, productivity, and reliability \cite{riek2025future}. These systems generate algorithmic operational decisions, but end users, the customers who receive the service, encounter those decisions through frontline workers who carry them out in real-world settings. In ride-hailing services such as Uber, for example, the system determines dispatch and routing, but passengers experience these decisions as service outcomes shaped through the driver's work, including waiting time, route, and arrival time. Users therefore encounter not the system's decision logic but outcomes that unfold through worker labor. These outcomes can diverge from user expectations. We define \textit{friction} as the user-facing tension that emerges when this divergence becomes salient during the service encounter. Frontline workers are commonly the first to face and manage such friction because the system remains in the background while the worker is present to the user. 

As automated systems enter the workplace, they reshape work practices and create new forms of labor \cite{suchman1995making}. CSCW scholarship on \textit{invisible labor} has examined human efforts that are essential for technological systems to function in practice but remain undervalued or unrecognized in formal accounts of work \cite{suchman1995making, star1999layers, riek2025future}. This scholarship shows how articulation work \cite{strauss1988articulation}, maintenance and repair \cite{orr1996talking, jackson2014rethinking}, and patchwork \cite{fox2023patchwork} sustain technical systems by addressing misalignments between formal procedures and situated realities. Such labor is often absorbed into routine operations and leaves little trace in performance metrics, system logs, or formal job descriptions. A prominent strand of this scholarship examines backstage labor that sustains automated systems away from end users \cite{gray2019ghost, irani2016hidden, alkhatib2017examining, roberts2019behind, wolf2020ai}, or becomes necessary when technical systems break down or fall short \cite{orr1996talking, suchman2007human, fox2023patchwork, cho2026thing}. A smaller body of work examines invisible labor in frontstage service encounters, where interaction with end users is itself the site of labor. Moradi et al. \cite{moradi2025pseudo}, for example, show how cashiers perform relational patchwork, including scapegoating machines and hyperservicing customers, to repair conflicts arising when self-checkout machines offset work onto customers. Building on this work, we examine a related but distinct frontstage condition in which automated systems generate operationally valid decisions and frontline workers mediate their consequences as they unfold against user expectations.

We refer to this labor as \textbf{Frontstage Mediation Work (FMW)}. FMW is the labor through which frontline workers anticipate and manage misalignments between automated decisions and end users' situated expectations at the service encounter. Following Goffman's distinction between frontstage and backstage regions of social performance \cite{goffman1959presentation}, we call this labor frontstage because automated decisions become consequential for users through the worker's visible service performance, while the system's decision logic remains backstage. The worker is a frontline worker by organizational position, whereas the mediation labor is frontstage because it is performed in the user's presence. Critically, FMW arises not only when a system malfunctions, but also when an operationally valid decision, produced as intended rather than through technical error, becomes experientially misaligned with user expectations. Because workers face users directly, they carry the interactional responsibility for these misalignments even though such mediation is rarely specified in their formal role.

To examine FMW in practice, we draw on an empirical case study of an AI-mediated On-Demand Ride-Pooling (ODRP) service. Unlike single-rider services such as Uber, ODRP pools multiple passengers into one vehicle and uses AI to dynamically generate operational decisions. These decisions reach passengers through drivers who carry them out, allowing misalignments between automated decisions and passenger expectations to surface directly at the service encounter. Through a qualitative analysis of ODRP drivers' work, we identify four recurring FMW practices through which drivers sustain encounters under such misalignments. FMW shifts attention from the point of decision, where automated decisions are generated, evaluated, or governed, to the point of encounter, where frontline workers absorb their consequences. We offer FMW as a preliminary analytic lens for examining how frontline workers make automated service decisions interactionally workable, and as a starting point for comparative study and theoretical development. 

\section{CASE STUDY: ON-DEMAND RIDE-POOLING SERVICE}
ODRP is a shared mobility service that coordinates multiple ride requests, dispatching one vehicle to pick up and drop off several passengers along a dynamically updated route. Core operational decisions, including ride matching, routing, stop sequencing, and arrival time estimates, are generated by an automated system integrating real-time demand, vehicle locations, and historical patterns \cite{alonso2017demand, zardini2022analysis, guo2022data}. Although these decisions are automated, the service depends on human drivers employed by the service operator to carry them out during service encounters. Passengers can also rate rides after each trip, and these ratings contribute to driver evaluation \cite{sohn2025driving}.

ODRP is well suited to examining FMW because its operational logic makes misalignments at the service encounter recurring rather than incidental. The system optimizes collective efficiency across multiple passengers rather than any individual ride, so operationally valid decisions can diverge from passengers' situated expectations regarding pickup and arrival times, routes, or drop-off sequence \cite{sohn2023exploring}. These misalignments surface face-to-face within the vehicle, where drivers and passengers share the same physical space. Whereas comparable tensions in food delivery or single-rider ride-hailing are often spatially distant or distributed across separate interfaces, ODRP concentrates them within a single co-present encounter. ODRP thus represents a concentrated ecology of automation. We approach it as an ultimate particular \cite{nelson2014design}, where the contexts, organizations, and workflows shaping automated services intersect in observable form. This particular ecology provides empirical grounding for examining frontstage mediation in future comparative studies. 

\section{METHOD}
Research Site: We conducted a qualitative case study of drivers at \textit{Shucle}, a South Korean ODRP service operated by Hyundai Motor Group. Launched in Sejong and Paju in 2021, it has since expanded nationwide \cite{SHUCLEREPRESENT2}. As one of the country's longest-running ODRP services, it provides a setting where both the system and drivers' practices have stabilized, making it suitable for examining FMW \cite{SHUCLEPOPULAR}. We had no financial or organizational ties to \textit{Shucle} or its operator. The interviews came from a broader study of ODRP drivers under algorithmic management \cite{sohn2025driving}; here, we analyze them specifically for FMW.

Participants and Data Collection: We recruited 16 drivers (Sejong: 5; Paju: 11) through on-site managers, bulletin boards, and snowball sampling. Participants were 12 men and 4 women (mean age=57.8 years, SD=6.0; mean tenure=15.3 months, SD=5.7). Data were collected 18--22 months after launch in each region, capturing practices that had developed alongside the service. We conducted 40--60-minute individual semi-structured interviews. Interviews began with questions on daily routines and overall work experiences, then followed issues raised by participants. Interviews were recorded and transcribed; participants received KRW 50,000 (approx. US\$37).

Data Analysis: We used exploratory inductive thematic analysis \cite{boyatzis1998transforming, braun2006using} to identify forms of labor in drivers' accounts rather than test predefined categories. We repeatedly reviewed transcripts and open-coded meaningful incidents, judgments, and interactions. We compared codes and grouped them into higher-level categories based on the conditions and contexts of related practices. This iterative process identified recurring practices through which drivers managed misalignments between automated decisions and passenger expectations at the service encounter. We call these practices FMW and present them next.

\section{FOUR PRACTICES OF FRONTSTAGE MEDIATION WORK IN ODRP SERVICE}
In ODRP, drivers carry out system decisions through real-world vehicle operation. While this execution is necessary for service delivery, it does not by itself ensure that passengers' expectations are met. As operationally valid decisions unfold in practice, misalignments can emerge between what the system decides and what passengers expect, calling for mediation work at the service encounter. Our analysis identifies four recurring practices of FMW, which differ in the kind of work they require: anticipating misalignment, absorbing passenger frustration, translating system decisions, and recalibrating passenger experience. A single situation can call for more than one of these practices. They are seldom formally recognized or specified, yet recur throughout drivers' work. 

\subsection{Anticipating Misalignment Before It Surfaces}
This practice refers to workers' cognitive labor of reading system decisions as they are carried out and judging whether they will create friction during the encounter. Although operationally valid, decisions may conflict with passengers' situated expectations about waiting time, routes, stop locations, or drop-off order. Drawing on experience and situated knowledge, workers assess whether a decision is likely to be accepted or trigger complaints or conflict. Rather than resolving friction, this practice diagnoses its risk before passengers voice it, allowing workers to prepare for friction that has not yet surfaced.

Drivers made these assessments even before pickup and throughout each ride. They considered how waiting times, stop points, drop-off order, and routing would be perceived, paying particular attention to unavoidable but potentially disadvantageous decisions such as detours, changes in estimated pickup or arrival times, and delayed drop-offs: ``\textit{When a detour appears on the screen, I already start watching the passenger's reaction through the rearview mirror, anticipating the friction to come}'' (P4). They also gauged the likely severity of each misalignment, treating substantial delays or detours as more likely to provoke complaints than minor route changes. Although not formally specified as part of their role, this cognitive work persisted alongside driving throughout the ride.

This practice is distinct from those that follow because it involves no outward action. Drivers do not yet apologize, explain, or change how they carry out the decision; they assess, in advance, whether it can be carried out without disrupting the encounter and whether further mediation may be required. This diagnostic work is visible neither to passengers nor to the system, yet it shapes every mediating move that follows, serving as the cognitive foundation for the other practices. 

\subsection{Absorbing Passenger Frustration to Keep the Encounter on Track}
This practice refers to workers' affective labor of managing passengers' emotional reactions when system decisions surface as friction during the service encounter. Because passengers experience decisions through the worker, complaints and frustration are often directed at the driver. While such emotional work is well established in service labor \cite{hochschild1979emotion}, here the affective burden falls on workers who must carry out automated decisions they did not make, and the system that produced them remains invisible to the passenger. In response, workers apologize, empathize, and adjust their demeanor to reduce tension and prevent escalation.

Detours and delayed arrival estimates, for example, could provoke negative reactions even when they resulted from operationally valid decisions. Despite having little control over these outcomes, drivers absorbed the resulting frustration while the system's rationale remained in the background: ``\textit{The ETA slips, and the passenger starts huffing, so I apologize like I'm the one who's late. I have to, or they'll just delete the app and never come back}'' (P12). This labor was not merely reactive. Drivers monitored the in-vehicle atmosphere and adjusted their tone and attitude throughout the ride to stabilize the encounter.

While this affective work helps maintain service continuity, it requires sustained emotion regulation and tolerance of negative interactions. It is rarely specified or recognized in formal accounts of the role, relying instead on drivers' personal skills and emotional resources. Drivers nonetheless treated it as pragmatically necessary; failing to absorb frustration would deteriorate the in-vehicle environment and compromise subsequent service delivery.

\subsection{Translating System Decisions into Situated Explanations}
This practice involves workers constructing situated explanations that help passengers interpret system decisions. Because the rationales underlying automated decisions are often unavailable during the encounter, passengers turn to drivers to explain what happened and why. Even without sufficient system access or information, workers draw on available cues, prior experience, and situated judgment to offer accounts that make decisions understandable in context. This interpretive work translates system-generated decisions into contextual accounts that relate system logic to passengers' situated expectations.

Drivers encountered questions about complex decisions such as routing, pickups, and drop-off sequences. Their formal role centered on executing these decisions, leaving them with limited training and access to the system's decision logic. They therefore constructed situated explanations in real time, taking on an interpretive burden to make opaque decisions intelligible. When a new pickup altered the route, for example, some explained that ``\textit{the route changed because another passenger joined}'' (P4), while others emphasized that ``\textit{this service pools multiple riders, so routes get adjusted}'' (P6) or that ``\textit{this is set as the most efficient route overall}'' (P12).

Providing such explanations was not formally required but difficult to avoid, as absent or unconvincing accounts could direct distrust toward drivers and provoke conflict. These explanations were provisional rather than technically complete. They provided situational meaning when the system remained opaque and helped sustain the encounter even though the driver had not produced the decision. Passengers saw the explanation itself, while the interpretive effort of reconstructing a plausible rationale without access to the system's logic remained unseen.

\subsection{Recalibrating Passenger Experience Through Situated Execution}
This practice involves adjusting the execution of system decisions in situ to recalibrate what passengers experience. Rather than altering the underlying decision, workers modify how it is carried out based on local conditions, passenger reactions, and the unfolding service context. By flexibly interpreting rules and adjusting the procedures, manner, or scope of execution, workers attempt to bring service outcomes closer to passengers' situated expectations. Unlike the anticipatory work in §4.1, which diagnoses risk without acting, this practice intervenes directly in how the decision is executed.

Drivers selectively modified routes or stop procedures when strict adherence was likely to provoke complaints or conflict. They used shortcuts or allowed drop-offs near designated stops in response to passenger requests. Such demands were difficult to ignore because drivers, rather than the system, bore their immediate emotional and interactional consequences. Some also intervened without explicit requests, taking personal risks to compensate for system-induced delays: ``\textit{When the delay is accumulated, I just start speeding to make up for the lost time}'' (P1).

Although these adjustments could reduce immediate tensions, they were neither consistently authorized nor legible at the organizational or system level. They responded not to technical breakdown but to operationally valid decisions that nonetheless surfaced as friction during the encounter. System logs might register route or stop deviations without capturing their situated rationale, exposing drivers to interpretations of noncompliance. Drivers therefore faced ongoing pressure to balance passenger demands against organizational rules. Although this practice produced the most observable forms of mediation, the situated judgment and context underlying it remained invisible to the system, leaving workers to manage misalignments without formal recognition or support.

\section{DISCUSSION AND FUTURE WORK}
Through an ODRP case, we propose FMW as an analytic lens for the labor through which workers mediate misalignments between automated decisions and passengers' situated expectations. FMW contributes to worker-centered HCI \cite{fox2020worker} by foregrounding the invisible labor that sustains service encounters when operationally valid decisions diverge from passenger expectations. Building on the articulation work tradition \cite{strauss1988articulation}, FMW specifies a frontstage configuration in which automated decisions are produced by the system, carried out by frontline workers, and assessed by passengers during the encounter. This configuration also places FMW in the dramaturgical tradition that understands service delivery as performance before an audience \cite{grove2013impression}. Workers manage both the driving task and passengers' responses to unwelcome outcomes, a form of what Goffman termed cooling the mark out \cite{goffman1952cooling, moradi2026goffmanian}.

FMW reveals a distinct visibility dynamic. Whereas much invisible-labor scholarship examines backstage efforts that sustain system functionality away from end users \cite{wolf2020ai, orr1996talking}, FMW is performed in frontstage encounters. Because work becomes visible through the indicators an audience can read \cite{star1999layers}, different audiences perceive FMW differently. In ODRP, system records register automated decisions but not the work of mediating them. The worker is visible as the accountable face of the service, while the mediation labor remains absent from metrics, system logs, and formal job descriptions. FMW thus involves a double asymmetry of visibility: the worker is interactionally salient, yet the labor of mediation disappears from institutional accounts. Anticipatory work (§4.1) is particularly invisible because neither passengers nor the organization observe the diagnostic judgment through which drivers assess the risk of friction before it surfaces.

FMW is not required for the system's technical operation, but it helps sustain the service encounter once automated decisions are carried out in front of passengers. Without it, friction remains unmanaged, and passengers may attribute outcomes to workers rather than to the underlying automation \cite{sohn2025driving}. Unlike breakdown repair or labor created by shifting tasks onto users \cite{moradi2025pseudo}, FMW addresses the interactional consequences of decisions that function as intended. It is therefore not formally mandated but practically difficult to avoid. Unlike contract-based ghost work \cite{gray2019ghost}, workers perform FMW because they bear the immediate interactional consequences of friction. In ODRP, passenger ratings further intensify this burden by attaching an evaluative record to the encounter and giving drivers a direct stake in dampening friction.

When FMW successfully prevents or reduces friction, it leaves little trace and is absorbed into routine operations. This creates conditions for unsupported mediation. Workers may construct explanations that diverge from system logic or selectively adjust how decisions are carried out, potentially compromising service consistency, accountability, and fairness. Its invisibility also limits system learning by obscuring where and why automated decisions diverge from passenger expectations. Making FMW visible, however, should not mean converting mediation into additional metrics or surveillance. Instead, automated services should reconsider how explanation, discretion, and accountability are distributed across systems, organizations, and frontline workers.

The implications of FMW extend beyond ODRP to service settings where automated decisions reach users through human labor. In food delivery, workers may anticipate customer complaints when delays emerge (§4.1) and deviate from assigned routes to reduce them (§4.4). In AI-driven financial services, frontline staff may manage applicants' emotional reactions to automated loan rejections (§4.2) and provide situated explanations for opaque decisions (§4.3). These cases suggest that FMW may arise wherever frontline workers carry out automated decisions that users evaluate as service experiences. As automation begins to remove the human mediator, as in autonomous ride-hailing or delivery robots, these interactional consequences do not disappear, raising the question of who, or what, will absorb them.

This study is grounded in a single service ecology and therefore has limited immediate transferability. We offer FMW as a preliminary analytic vocabulary rather than a finalized concept. Future work should examine FMW across healthcare, finance, public services, and delivery, and investigate how organizational norms, accountability structures, and system design shape its forms and visibility. It should also trace the temporal relations among the four practices, including how anticipatory work (§4.1) shapes responsive practices (§4.2 to §4.4) over the course of an encounter. Such studies can further clarify the structural conditions under which human mediation becomes invisible in automated service environments.

\section{CONCLUSION}
We propose FMW as a preliminary analytic lens for the invisible labor through which frontline workers mediate misalignments between automated decisions and users' situated expectations. Grounded in an ODRP case, we identify four recurring practices that sustain service encounters when operationally valid decisions diverge from passenger expectations. Because this labor is absorbed into routine operations and remains formally unrecognized, responsibility for the interactional consequences of automated decisions shifts onto frontline workers. FMW shows that automated service continuity can depend on frontstage human mediation and calls on service providers to recognize, support, and set boundaries around this labor.

\begin{acks}
We extend our gratitude to all participants in our study. We also sincerely thank the anonymous reviewers for their insightful comments and suggestions. This work was supported by the National Research Foundation of Korea (NRF) grants funded by the Korea government (MSIT) (RS-2026-25471667 and No. NRF-2021R1A2C2004263).
\end{acks}

\bibliographystyle{ACM-Reference-Format}
\bibliography{sample-base}


\begin{thebibliography}{31}

%%% the Web address.

\ifx \showCODEN    \undefined \def \showCODEN     #1{\unskip}     \fi
\ifx \showISBNx    \undefined \def \showISBNx     #1{\unskip}     \fi
\ifx \showISBNxiii \undefined \def \showISBNxiii  #1{\unskip}     \fi
\ifx \showISSN     \undefined \def \showISSN      #1{\unskip}     \fi
\ifx \showLCCN     \undefined \def \showLCCN      #1{\unskip}     \fi
\ifx \shownote     \undefined \def \shownote      #1{#1}          \fi
\ifx \showarticletitle \undefined \def \showarticletitle #1{#1}   \fi
\ifx \showURL      \undefined \def \showURL       {\relax}        \fi
% invisible to TeX
\providecommand\bibfield[2]{#2}
\providecommand\bibinfo[2]{#2}
\providecommand\natexlab[1]{#1}
\providecommand\showeprint[2][]{arXiv:#2}

\bibitem[Alkhatib et~al\mbox{.}(2017)]%
        {alkhatib2017examining}
\bibfield{author}{\bibinfo{person}{Ali Alkhatib}, \bibinfo{person}{Michael~S Bernstein}, {and} \bibinfo{person}{Margaret Levi}.} \bibinfo{year}{2017}\natexlab{}.
\newblock \showarticletitle{Examining crowd work and gig work through the historical lens of piecework}. In \bibinfo{booktitle}{\emph{Proceedings of the 2017 CHI conference on human factors in computing systems}}. \bibinfo{pages}{4599--4616}.
\newblock


\bibitem[Alonso-Mora et~al\mbox{.}(2017)]%
        {alonso2017demand}
\bibfield{author}{\bibinfo{person}{Javier Alonso-Mora}, \bibinfo{person}{Samitha Samaranayake}, \bibinfo{person}{Alex Wallar}, \bibinfo{person}{Emilio Frazzoli}, {and} \bibinfo{person}{Daniela Rus}.} \bibinfo{year}{2017}\natexlab{}.
\newblock \showarticletitle{On-demand high-capacity ride-sharing via dynamic trip-vehicle assignment}.
\newblock \bibinfo{journal}{\emph{Proceedings of the National Academy of Sciences}} \bibinfo{volume}{114}, \bibinfo{number}{3} (\bibinfo{year}{2017}), \bibinfo{pages}{462--467}.
\newblock


\bibitem[Boyatzis(1998)]%
        {boyatzis1998transforming}
\bibfield{author}{\bibinfo{person}{Richard~E Boyatzis}.} \bibinfo{year}{1998}\natexlab{}.
\newblock \bibinfo{booktitle}{\emph{Transforming qualitative information: Thematic analysis and code development}}.
\newblock \bibinfo{publisher}{sage}.
\newblock


\bibitem[Braun and Clarke(2006)]%
        {braun2006using}
\bibfield{author}{\bibinfo{person}{Virginia Braun} {and} \bibinfo{person}{Victoria Clarke}.} \bibinfo{year}{2006}\natexlab{}.
\newblock \showarticletitle{Using thematic analysis in psychology}.
\newblock \bibinfo{journal}{\emph{Qualitative research in psychology}} \bibinfo{volume}{3}, \bibinfo{number}{2} (\bibinfo{year}{2006}), \bibinfo{pages}{77--101}.
\newblock


\bibitem[Cho et~al\mbox{.}(2026)]%
        {cho2026thing}
\bibfield{author}{\bibinfo{person}{Hyungjun Cho}, \bibinfo{person}{Jiyeon~Amy Seo}, \bibinfo{person}{Yongjae Sohn}, {and} \bibinfo{person}{Hee~Rin Lee}.} \bibinfo{year}{2026}\natexlab{}.
\newblock \showarticletitle{Thing Ethnography in a Factory: Exploring Emergent and Dynamic Relations of Cobots and Workers}. In \bibinfo{booktitle}{\emph{Proceedings of the 2026 CHI Conference on Human Factors in Computing Systems}}. \bibinfo{pages}{1--21}.
\newblock


\bibitem[Fox et~al\mbox{.}(2020)]%
        {fox2020worker}
\bibfield{author}{\bibinfo{person}{Sarah~E Fox}, \bibinfo{person}{Vera Khovanskaya}, \bibinfo{person}{Clara Crivellaro}, \bibinfo{person}{Niloufar Salehi}, \bibinfo{person}{Lynn Dombrowski}, \bibinfo{person}{Chinmay Kulkarni}, \bibinfo{person}{Lilly Irani}, {and} \bibinfo{person}{Jodi Forlizzi}.} \bibinfo{year}{2020}\natexlab{}.
\newblock \showarticletitle{Worker-centered design: Expanding HCI methods for supporting labor}. In \bibinfo{booktitle}{\emph{Extended abstracts of the 2020 CHI conference on human factors in computing systems}}. \bibinfo{pages}{1--8}.
\newblock


\bibitem[Fox et~al\mbox{.}(2023)]%
        {fox2023patchwork}
\bibfield{author}{\bibinfo{person}{Sarah~E Fox}, \bibinfo{person}{Samantha Shorey}, \bibinfo{person}{Esther~Y Kang}, \bibinfo{person}{Dominique Montiel~Valle}, {and} \bibinfo{person}{Estefania Rodriguez}.} \bibinfo{year}{2023}\natexlab{}.
\newblock \showarticletitle{Patchwork: the hidden, human labor of AI integration within essential work}.
\newblock \bibinfo{journal}{\emph{Proceedings of the ACM on Human-Computer Interaction}} \bibinfo{volume}{7}, \bibinfo{number}{CSCW1} (\bibinfo{year}{2023}), \bibinfo{pages}{1--20}.
\newblock


\bibitem[Goffman(1952)]%
        {goffman1952cooling}
\bibfield{author}{\bibinfo{person}{Erving Goffman}.} \bibinfo{year}{1952}\natexlab{}.
\newblock \showarticletitle{On cooling the mark out: Some aspects of adaptation to failure}.
\newblock \bibinfo{journal}{\emph{Psychiatry}} \bibinfo{volume}{15}, \bibinfo{number}{4} (\bibinfo{year}{1952}), \bibinfo{pages}{451--463}.
\newblock


\bibitem[Goffman(1959)]%
        {goffman1959presentation}
\bibfield{author}{\bibinfo{person}{Erving Goffman}.} \bibinfo{year}{1959}\natexlab{}.
\newblock \bibinfo{booktitle}{\emph{The Presentation of Self in Everyday Life}}.
\newblock \bibinfo{publisher}{Doubleday Anchor}, \bibinfo{address}{Garden City, NY}.
\newblock


\bibitem[Gray and Suri(2019)]%
        {gray2019ghost}
\bibfield{author}{\bibinfo{person}{Mary~L Gray} {and} \bibinfo{person}{Siddharth Suri}.} \bibinfo{year}{2019}\natexlab{}.
\newblock \bibinfo{booktitle}{\emph{Ghost Work: How to Stop Silicon Valley from Building a New Global Underclass}}.
\newblock \bibinfo{publisher}{Houghton Mifflin Harcourt}, \bibinfo{address}{Boston, MA}.
\newblock


\bibitem[Grove and Fisk(2013)]%
        {grove2013impression}
\bibfield{author}{\bibinfo{person}{Stephen~J Grove} {and} \bibinfo{person}{Raymond~P Fisk}.} \bibinfo{year}{2013}\natexlab{}.
\newblock \showarticletitle{Impression management in services marketing: A dramaturgical perspective}.
\newblock In \bibinfo{booktitle}{\emph{Impression management in the organization}}. \bibinfo{publisher}{Psychology Press}, \bibinfo{pages}{427--438}.
\newblock


\bibitem[Guo et~al\mbox{.}(2022)]%
        {guo2022data}
\bibfield{author}{\bibinfo{person}{Xiaotong Guo}, \bibinfo{person}{Qingyi Wang}, {and} \bibinfo{person}{Jinhua Zhao}.} \bibinfo{year}{2022}\natexlab{}.
\newblock \showarticletitle{Data-driven vehicle rebalancing with predictive prescriptions in the ride-hailing system}.
\newblock \bibinfo{journal}{\emph{IEEE Open Journal of Intelligent Transportation Systems}}  \bibinfo{volume}{3} (\bibinfo{year}{2022}), \bibinfo{pages}{251--266}.
\newblock


\bibitem[Hochschild(1979)]%
        {hochschild1979emotion}
\bibfield{author}{\bibinfo{person}{Arlie~Russell Hochschild}.} \bibinfo{year}{1979}\natexlab{}.
\newblock \showarticletitle{Emotion work, feeling rules, and social structure}.
\newblock \bibinfo{journal}{\emph{American journal of sociology}} \bibinfo{volume}{85}, \bibinfo{number}{3} (\bibinfo{year}{1979}), \bibinfo{pages}{551--575}.
\newblock


\bibitem[Irani(2016)]%
        {irani2016hidden}
\bibfield{author}{\bibinfo{person}{Lilly Irani}.} \bibinfo{year}{2016}\natexlab{}.
\newblock \showarticletitle{The hidden faces of automation}.
\newblock \bibinfo{journal}{\emph{XRDS: Crossroads, The ACM Magazine for Students}} \bibinfo{volume}{23}, \bibinfo{number}{2} (\bibinfo{year}{2016}), \bibinfo{pages}{34--37}.
\newblock


\bibitem[Jackson(2014)]%
        {jackson2014rethinking}
\bibfield{author}{\bibinfo{person}{Steven~J Jackson}.} \bibinfo{year}{2014}\natexlab{}.
\newblock \showarticletitle{Rethinking Repair}.
\newblock In \bibinfo{booktitle}{\emph{Media Technologies: Essays on Communication, Materiality, and Society}}, \bibfield{editor}{\bibinfo{person}{Tarleton Gillespie}, \bibinfo{person}{Pablo~J Boczkowski}, {and} \bibinfo{person}{Kirsten~A Foot}} (Eds.). \bibinfo{publisher}{MIT Press}, \bibinfo{address}{Cambridge, MA}, \bibinfo{pages}{221--239}.
\newblock


\bibitem[Moradi et~al\mbox{.}(2025)]%
        {moradi2025pseudo}
\bibfield{author}{\bibinfo{person}{Pegah Moradi}, \bibinfo{person}{Karen Levy}, {and} \bibinfo{person}{Cristobal Cheyre}.} \bibinfo{year}{2025}\natexlab{}.
\newblock \showarticletitle{Pseudo-automation: how labor-offsetting technologies reconfigure roles and relationships in frontline retail work}.
\newblock \bibinfo{journal}{\emph{Proceedings of the ACM on Human-Computer Interaction}} \bibinfo{volume}{9}, \bibinfo{number}{2} (\bibinfo{year}{2025}), \bibinfo{pages}{1--21}.
\newblock


\bibitem[Moradi et~al\mbox{.}(2026)]%
        {moradi2026goffmanian}
\bibfield{author}{\bibinfo{person}{Pegah Moradi}, \bibinfo{person}{Karen Levy}, {and} \bibinfo{person}{Elizabeth Chiarello}.} \bibinfo{year}{2026}\natexlab{}.
\newblock \showarticletitle{Goffmanian “cooling” in technology-mediated frontline enforcement work}.
\newblock \bibinfo{journal}{\emph{Social Science Computer Review}} \bibinfo{volume}{44}, \bibinfo{number}{1} (\bibinfo{year}{2026}), \bibinfo{pages}{51--58}.
\newblock


\bibitem[Nelson and Stolterman(2014)]%
        {nelson2014design}
\bibfield{author}{\bibinfo{person}{Harold~G Nelson} {and} \bibinfo{person}{Erik Stolterman}.} \bibinfo{year}{2014}\natexlab{}.
\newblock \bibinfo{booktitle}{\emph{The design way: Intentional change in an unpredictable world}}.
\newblock \bibinfo{publisher}{MIT press}.
\newblock


\bibitem[Orr(1996)]%
        {orr1996talking}
\bibfield{author}{\bibinfo{person}{Julian~E Orr}.} \bibinfo{year}{1996}\natexlab{}.
\newblock \bibinfo{booktitle}{\emph{Talking about Machines: An Ethnography of a Modern Job}}.
\newblock \bibinfo{publisher}{Cornell University Press}, \bibinfo{address}{Ithaca, NY}.
\newblock


\bibitem[Riek and Irani(2025)]%
        {riek2025future}
\bibfield{author}{\bibinfo{person}{Laurel~D Riek} {and} \bibinfo{person}{Lilly Irani}.} \bibinfo{year}{2025}\natexlab{}.
\newblock \showarticletitle{The Future Is Rosie?: Disempowering Arguments About Automation and What to Do About It}. In \bibinfo{booktitle}{\emph{Proceedings of the 2025 CHI Conference on Human Factors in Computing Systems}}. \bibinfo{pages}{1--14}.
\newblock


\bibitem[Roberts(2019)]%
        {roberts2019behind}
\bibfield{author}{\bibinfo{person}{Sarah~T Roberts}.} \bibinfo{year}{2019}\natexlab{}.
\newblock \bibinfo{booktitle}{\emph{Behind the screen}}.
\newblock \bibinfo{publisher}{Yale University Press}.
\newblock


\bibitem[Shucle(2023)]%
        {SHUCLEREPRESENT2}
\bibfield{author}{\bibinfo{person}{Shucle}.} \bibinfo{year}{2023}\natexlab{}.
\newblock \bibinfo{booktitle}{\emph{Shucle}}.
\newblock
\urldef\tempurl%
\url{https://www.shucle.com/}
\showURL{%
\tempurl}


\bibitem[Shucle(2024)]%
        {SHUCLEPOPULAR}
\bibfield{author}{\bibinfo{person}{Shucle}.} \bibinfo{year}{2024}\natexlab{}.
\newblock \bibinfo{booktitle}{\emph{Hyundai Shucle's 2024, a helping hand for the transportation disadvantaged}}.
\newblock
\urldef\tempurl%
\url{https://www.hyundai.co.kr/story/CONT0000000000167183}
\showURL{%
\tempurl}


\bibitem[Sohn et~al\mbox{.}(2025)]%
        {sohn2025driving}
\bibfield{author}{\bibinfo{person}{Yongjae Sohn}, \bibinfo{person}{Daehyun Kwak}, \bibinfo{person}{Sehee Son}, \bibinfo{person}{Chowon Kang}, {and} \bibinfo{person}{Youn-kyung Lim}.} \bibinfo{year}{2025}\natexlab{}.
\newblock \showarticletitle{Driving with Algorithms Beyond Gig Work: Investigating How Algorithmic Management Affects Workers’ Practices in an On-Demand Ride-Pooling Service}. In \bibinfo{booktitle}{\emph{Proceedings of the 2025 ACM Designing Interactive Systems Conference}}. \bibinfo{pages}{2117--2129}.
\newblock


\bibitem[Sohn et~al\mbox{.}(2023)]%
        {sohn2023exploring}
\bibfield{author}{\bibinfo{person}{Yongjae Sohn}, \bibinfo{person}{Sehee Son}, {and} \bibinfo{person}{Youn-Kyung Lim}.} \bibinfo{year}{2023}\natexlab{}.
\newblock \showarticletitle{Exploring passengers’ experiences in algorithmic dynamic ride-pooling services}. In \bibinfo{booktitle}{\emph{Companion Publication of the 2023 ACM Designing Interactive Systems Conference}}. \bibinfo{pages}{237--240}.
\newblock


\bibitem[Star and Strauss(1999)]%
        {star1999layers}
\bibfield{author}{\bibinfo{person}{Susan~Leigh Star} {and} \bibinfo{person}{Anselm Strauss}.} \bibinfo{year}{1999}\natexlab{}.
\newblock \showarticletitle{Layers of silence, arenas of voice: The ecology of visible and invisible work}.
\newblock \bibinfo{journal}{\emph{Computer supported cooperative work (CSCW)}} \bibinfo{volume}{8}, \bibinfo{number}{1} (\bibinfo{year}{1999}), \bibinfo{pages}{9--30}.
\newblock


\bibitem[Strauss(1988)]%
        {strauss1988articulation}
\bibfield{author}{\bibinfo{person}{Anselm Strauss}.} \bibinfo{year}{1988}\natexlab{}.
\newblock \showarticletitle{The articulation of project work: An organizational process}.
\newblock \bibinfo{journal}{\emph{Sociological Quarterly}} \bibinfo{volume}{29}, \bibinfo{number}{2} (\bibinfo{year}{1988}), \bibinfo{pages}{163--178}.
\newblock


\bibitem[Suchman(1995)]%
        {suchman1995making}
\bibfield{author}{\bibinfo{person}{Lucy Suchman}.} \bibinfo{year}{1995}\natexlab{}.
\newblock \showarticletitle{Making work visible}.
\newblock \bibinfo{journal}{\emph{Commun. ACM}} \bibinfo{volume}{38}, \bibinfo{number}{9} (\bibinfo{year}{1995}), \bibinfo{pages}{56--64}.
\newblock


\bibitem[Suchman(2007)]%
        {suchman2007human}
\bibfield{author}{\bibinfo{person}{Lucille~Alice Suchman}.} \bibinfo{year}{2007}\natexlab{}.
\newblock \bibinfo{booktitle}{\emph{Human-machine reconfigurations: Plans and situated actions}}.
\newblock \bibinfo{publisher}{Cambridge university press}.
\newblock


\bibitem[Wolf(2020)]%
        {wolf2020ai}
\bibfield{author}{\bibinfo{person}{Christine~T Wolf}.} \bibinfo{year}{2020}\natexlab{}.
\newblock \showarticletitle{AI models and their worlds: Investigating data-driven, AI/ML ecosystems through a work practices lens}. In \bibinfo{booktitle}{\emph{International conference on information}}. Springer, \bibinfo{pages}{651--664}.
\newblock


\bibitem[Zardini et~al\mbox{.}(2022)]%
        {zardini2022analysis}
\bibfield{author}{\bibinfo{person}{Gioele Zardini}, \bibinfo{person}{Nicolas Lanzetti}, \bibinfo{person}{Marco Pavone}, {and} \bibinfo{person}{Emilio Frazzoli}.} \bibinfo{year}{2022}\natexlab{}.
\newblock \showarticletitle{Analysis and control of autonomous mobility-on-demand systems}.
\newblock \bibinfo{journal}{\emph{Annual Review of Control, Robotics, and Autonomous Systems}} \bibinfo{volume}{5}, \bibinfo{number}{1} (\bibinfo{year}{2022}), \bibinfo{pages}{633--658}.
\newblock


\end{thebibliography}
\end{document}